\documentclass[prc,twocolumn,nofootinbib]{revtex4}

\usepackage{graphicx}
\usepackage[usenames,dvipsnames]{color}
\usepackage{amsmath,amssymb,bbold}
\usepackage{hyperref}
\usepackage{comment}
\usepackage{appendix}
\date{\today}

\begin{document}

\title{Correspondence between Israel-Stewart and first-order causal and stable hydrodynamics for Bjorken-expanding baryon-rich systems with vanishing particle masses}

\author{Arpan Das$^{1}$}
\email{arpan.das@ifj.edu.pl}
\author{Wojciech Florkowski$^{2}$}
\email{wojciech.florkowski@uj.edu.pl}
\author{Radoslaw Ryblewski${^1}$}
\email{radoslaw.ryblewski@ifj.edu.pl}

\affiliation{$^{1}$Institute  of  Nuclear  Physics  Polish  Academy  of  Sciences,  PL-31-342  Krak\'ow,  Poland}

\affiliation{$^{2}$Institute of Theoretical Physics, Jagiellonian University, PL-30-348 Krak\'ow, Poland}

\begin{abstract}
We obtain an exact correspondence between the dynamical equations in Israel-Stewart (IS) theory and first-order causal and stable (FOCS) hydrodynamics for a boost-invariant system with an ideal gas equation of state at finite baryon chemical potential. Explicit expressions for the temperature and chemical potential dependence of the regulators in the FOCS theory are given in terms of the kinetic coefficients and constant relaxation time of the IS theory. Using the correspondence between the IS and FOCS theory, stability conditions for a charged fluid which are known in the FOCS approach are applied and one finds that the IS theory considered is unstable.  
\end{abstract}

\maketitle

\section{Introduction}
Relativistic hydrodynamics has been proved to be very successful in describing the bulk evolution of the strongly interacting matter produced in heavy-ion collision experiments~\cite{Florkowski:2010zz,Gale:2013da,Jeon:2015dfa, Jaiswal:2016hex,Busza:2018rrf}. In connection with this success, various formal aspects of relativistic hydrodynamics have been extensively explored~\cite{Romatschke:2017ejr,Florkowski:2017olj}. Along with the modeling of the early stages of heavy-ion collisions and late-stage freeze-out of hadrons, relativistic hydrodynamics constitutes the standard model of heavy-ion collision experiments.

The formalism of relativistic hydrodynamics  includes many interesting new features such as
the existence of hydrodynamic attractors~\cite{Heller:2015dha,Romatschke:2017vte,Strickland:2017kux,Strickland:2018ayk,Jaiswal:2019cju,Giacalone:2019ldn} and asymptotic character of the hydrodynamic gradient
expansion~\cite{Heller:2013fn,Denicol:2016bjh,Heller:2016rtz,Grozdanov:2019kge}. Various properties of strongly interacting matter, e.g., equation of state,  bulk viscosity, shear viscosity, etc., can be obtained by a detailed comparison between theoretical predictions based on the hydrodynamic approach and available experimental data~\cite{Broniowski:2008vp,Bozek:2009dw,Noronha-Hostler:2013gga,Bernhard:2019bmu,Romatschke:2007mq,Gale:2012rq}. 

Theory of first-order viscous hydrodynamics dates back to 1950's when Landau and Eckart came up with a relativistic generalization of the non-relativistic Navier-Stokes equations. However, these theories are known to be in contradiction with causality and stability requirements, features that are very important for any acceptable theory of relativistic hydrodynamics~\cite{Hiscock:1983zz,Hiscock:1985zz}. Therefore, the first-order Landau and Eckart hydrodynamics have been discarded and replaced by a more general second-order Israel-Stewart (IS) theory~\cite{Israel:1976tn,Israel:1979wp}. The IS approach has been extensively used to describe the space-time evolution of the strongly interacting matter produced at the Relativistic Heavy Ion Collider (RHIC) at Brookheaven National Laboratory and the Large Hadron Collider (LHC) at CERN.

In the IS theory, the shear stress tensor ($\pi^{\mu\nu}$),  the bulk pressure ($\Pi$), and the particle diffusion (baryon diffusion) current ($n^{\mu}$) are considered as independent hydrodynamic variables, similar to  the local temperature ($T$), chemical potential ($\mu$), and the fluid four-velocity ($u^{\mu}$). They
approach their Navier-Stokes values, i.e., $\pi^{\mu\nu}=2\eta\sigma^{\mu\nu}$, $\Pi=-\zeta\partial_{\mu}u^{\mu}$ and $n^{\mu}=\kappa_n\Delta^{\mu\alpha}\partial_{\alpha}(\mu/T)$ during the hydrodynamic evolution, when the system approaches local equilibrium. Here $\eta$ and $\zeta$ are the coefficients of the shear and bulk viscosity, respectively, $\kappa_n$ is the charge conductivity coefficient that can also be related to the thermal conductivity,  $\sigma^{\mu\nu}$ is the shear flow tensor constructed from the derivatives of $u^{\mu}$, and $\Delta^{\mu}_{\nu}$ is the projection operator orthogonal to the fluid four-velocity ($u^{\mu}$).

Very recently the formulation of the first-order causal and stable (FOCS) relativistic viscous hydrodynamics has been put forward by Bemfica, Disconzi, Noronha, and Kovtun ~\cite{Bemfica:2017wps,Bemfica:2019knx,Kovtun:2019hdm}. In the FOCS formalism, one deals only with the Navier-Stokes degrees of freedom, i.e., local temperature ($T$), chemical potential ($\mu$), and the fluid four-velocity ($u^{\mu}$) as fundamental hydrodynamic variables in the derivative expansion. Rather than choosing some specific fluid frame, e.g., the Landau frame or the Eckart frame, FOCS is based on a more general choice of hydrodynamic frames and, relatedly, the introduction of a new set of kinetic coefficients. These new kinetic coefficients act as an ultraviolet regulators of the theory and make it causal (even in the fully nonlinear regime) and linearly stable around equilibrium.

Given various causal and stable theories of viscous relativistic hydrodynamics, IS and FOCS theories in particular, one can naturally ask for any possible correspondence between these theories. For a boost-invariant system with conformal symmetry these two approaches, i.e., IS and FOCS, lead to very similar equations~\cite{Bemfica:2017wps}. In our previous papers~\cite{Das:2020fnr,Das:2020gtq}, we have found that there is an exact matching between FOCS and IS equations for boost-invariant, Bjorken expanding systems with massless as well as massive particles for vanishing baryon number density. Due to the exact mapping between IS and FOCS theories, we could use the knowledge of non-linear causality and stability in the FOCS theory to study causality and the stability of IS theory. In the present investigation, we extend our previous works to a fluid with conserved charge (here we consider only baryon number conservation). 

It is important to note that on general grounds IS and FOCS theories are quite different and no direct connection between these two theories may exist. This is so because IS theory leads to ten first order differential equations, while FOCS gives four second order differential equations or equivalently eight first-order differential equations. Therefore, only for some special cases an exact correspondence between these two theories may exist (where the two frameworks lead to the same dynamical equations). Finding such correspondence is important for theoretical as well as phenomenological reasons. Using a mapping between IS and FOCS, we can transfer the knowledge gained in one sector to the other one. For example, non-linear causality and stability in the IS theory has not been explored for a long time --- only recently the non-linear causality and stability of an uncharged fluid has been investigated in Ref.~\cite{Bemfica:2020xym}. On the other hand, the causality and the stability of the FOCS theory is known~\cite{Bemfica:2017wps,Bemfica:2019knx,Kovtun:2019hdm,Hoult:2020eho}. Therefore, the information about causality and stability established in the FOCS approach can be used to analyze the causality and stability of IS solutions, provided such correspondence exists.

The paper is organized in the following manner: In Sec.~\ref{sec2} we introduce the IS and FOCS hydrodynamic equations for a fluid with conserved charge (here we consider baryon number conservation). In Sec.~\ref{sec3} we present a matching between these two frameworks using the approximation of constant relaxation time. The correspondence between the IS and FOCS theories is then used to discuss the stability of the IS theory in Sec.~\ref{caisality_stability}. We summarize and conclude in Sec.~\ref{summary}. Throughout the paper we use natural units $\hbar =c = k_B=1$.


\section{IS and FOCS hydrodynamics}
\label{sec2}
\subsection{IS approach}

 We start with the IS theory as given in Refs.~\cite{Jaiswal:2013fc,Jaiswal:2015mxa} to seek for a correspondence between the IS and FOCS theories in the case of nonvanishing baryon chemical potential $(\mu)$ and for an ideal-gas equation of state. The energy-momentum tensor and the conserved baryon number current for massless particles, as given in Refs.~\cite{Jaiswal:2013fc,Jaiswal:2015mxa}, are
\begin{align}
T^{\mu\nu} & = \varepsilon u^\mu u^\nu-p\Delta ^{\mu \nu} 
+ \pi^{\mu\nu},\label{equ1}\\
N^\mu & = nu^\mu + n^\mu \label{equ2}.
\end{align}
In Eqs.~\eqref{equ1} and \eqref{equ2}, the quantities $\varepsilon$, $p$, $n$, and $u^{\mu}$ represent energy density, pressure, number density, and fluid four-velocity, respectively. The quantities $\pi^{\mu\nu}$ (shear stress tensor) and $n^{\mu}$ (baryon diffusion current) are the corresponding dissipative parts of the energy-momentum tensor and the number current, respectively. 

Conservation of the energy-momentum tensor, $\partial_\mu T^{\mu\nu} =0$, and the baryon current conservation, $\partial_\mu N^{\mu}=0$, yield the evolution equations of energy density ($\varepsilon$), fluid four-velocity ($u^\mu$), and the baryon or number density ($n$)
\cite{Jaiswal:2013fc,Jaiswal:2015mxa},
\begin{align}
\dot\varepsilon + (\varepsilon+p)\partial_\mu u^\mu - \pi^{\mu\nu}\sigma_{\mu\nu} &= 0,  \nonumber\\
(\varepsilon+p)\dot u^\alpha - \nabla^\alpha p + \Delta^\alpha_\nu \partial_\mu \pi^{\mu\nu}  &= 0,  \nonumber\\
\dot n + n\partial_\mu u^\mu + \partial_\mu n^{\mu} &=0.
\label{equ3}
\end{align}
Here we have used the notation $\dot{A}=u^{\mu}\partial_{\mu}A$ for comoving derivative, $\nabla^\mu=\Delta^{\mu\nu}\partial_{\nu}$, and $\Delta^{\mu\nu}$  is the projector orthogonal to the fluid four velocity $u^{\mu}$.
To close the system of hydrodynamic equations we also need the evolution equations for the dissipative fluxes ($\pi^{\mu\nu}$, $n^\mu$) along with the equation of state (in our case $\varepsilon=3p$). Using relativistic kinetic theory, the evolution equations of these dissipative currents have been obtained in Refs.~\cite{Jaiswal:2013fc,Jaiswal:2015mxa}. They have the following form
\begin{align}
&\dot\pi^{\langle\mu\nu\rangle} + \frac{\pi^{\mu\nu}}{\tau_\pi} =
2\beta_\pi\sigma^{\mu\nu}
+ 2\pi_\gamma^{\langle\mu}\omega^{\nu\rangle\gamma}
-\frac{4}{3}\pi^{\mu\nu}\theta\nonumber\\
&~~~~~~~~~~~~~~~~~~~~~~~~~~~~~~~~~~~~~~~~~-\frac{10}{7}\pi_\gamma^{\langle\mu}\sigma^{\nu\rangle\gamma}, \label{equ4}\\
& \dot n^{\langle\mu\rangle} + \frac{n^\mu}{\tau_n} =
\beta_n\nabla^\mu\alpha
-n_\nu\omega^{\nu\mu}
-n^\mu\theta
-\frac{3}{5}n_\nu\sigma^{\nu\mu}\nonumber\\
& ~~~~~~~~~~~~~~~~~~~~~~~~~~~~~~~~~~~~~~~~-\frac{3\beta_n}{\varepsilon+p}\pi^{\mu\nu}\nabla_\nu\alpha.
\label{equ5}
\end{align}
Here $\omega^{\mu\nu}\equiv(\nabla^\mu u^\nu-\nabla^\nu u^\mu)/2$ is the anti-symmetric vorticity tensor, $\sigma^{\mu\nu} \equiv \frac{1}{2}(\nabla^{\mu}u^{\nu}+\nabla^{\nu}u^{\mu}
)-\frac{1}{3}\theta \Delta^{\mu\nu}$, $\theta = \partial_{\mu}u^{\mu}$ is the expansion scalar, $\alpha=\mu/T$, $\tau_\pi=\eta/\beta_\pi$, and 
$\tau_n=\kappa_n/\beta_n$ \cite{Jaiswal:2013fc,Jaiswal:2015mxa}. The quantity $\eta$ is the shear viscosity coefficient, $\mu$ and $T$ denote baryon chemical potential and temperature respectively. 

For simplicity, let us first assume that the baryon diffusion current vanishes, $n^{\mu}=0$ (the case where $n^{\mu}\neq0$ is discussed below). For the Bjorken flow the hydrodynamic equations given by Eqs.~\eqref{equ3} and the evolution equations for the dissipative quantities given by Eqs.~\eqref{equ4} and \eqref{equ5} simplify in this case to the following set of equations~\cite{Jaiswal:2013fc,Jaiswal:2015mxa}:
  \begin{align}
  & \frac{d\varepsilon}{d\tau} = -\frac{1}{\tau} \bigg[(\varepsilon+p)-\pi\bigg],\label{equ6}\\
  & \tau_{\pi}\frac{d\pi}{d\tau}= \frac{4}{3}\frac{\eta}{\tau} -\pi -\beta \frac{\tau_{\pi}}{\tau}\pi,
  \label{equ7}\\
  & \frac{dn}{d\tau}+\frac{n}{\tau}=0.
  \label{equ8}
 \end{align}
 Here $\tau \equiv\sqrt{t^2-z^2}$ is the longitudinal proper time and $\pi\equiv \pi^{\eta_s\eta_s}$ is the rapidity-rapidity component of the shear stress tensor $\pi^{\mu\nu}$ (representation of $\pi^{\mu\nu}$ for the Bjorken symmetry in the Milne coordinates is given in Ref.~\cite{Denicol:2017lxn}). In Eq.~\eqref{equ7} one uses $\beta = 4/3+\lambda$ with $\lambda=10/21$ \cite{Jaiswal:2013fc,Jaiswal:2015mxa}. In this work, however, we consider $\beta$ to be a free parameter. 
 
 It can be shown that even for a non vanishing baryon diffusion current the boost-invariant versions of Eqs.~\eqref{equ6}--\eqref{equ8} remain unchanged~\footnote{ For the Bjorken flow, in the Milne coordinates $n^{\mu}\equiv (0,0,0,n^{\eta_s})$ and $u^{\mu}\equiv (1,0,0,0)$~\cite{Du:2019obx}. Note that all the dissipative currents are orthogonal to $u^{\mu}$. Therefore, in the Milne coordinates the conservation of the baryon current implies $D_{\mu}N^{\mu}=u^{\mu}D_{\mu}n+nD_{\mu}u^{\mu}+D_{\mu}n^{\mu}=0$. Here $D_{\mu}$ is the covariant derivative. It can be easily shown that $D_{\mu}n^{\mu}=0$ for the Bjorken flow. Therefore one obtains $D_{\mu}N^{\mu}=\frac{dn}{d\tau}+\frac{n}{\tau}=0$ for the Bjorken flow even for nonvanishing $n^{\mu}$.} --- in this case we have an additional evolution equation for $n^{\mu}$ that completely decouples from the rest of~Eqs.~\eqref{equ6}--\eqref{equ8}~\cite{Du:2019obx}. Therefore, even for $n^{\mu}\neq 0$ we can proceed with Eqs.~\eqref{equ6}--\eqref{equ8} to match the FOCS boost-invariant formulation. In the following we will use the approximation of constant relaxation time in the shear sector with $\tau_{\pi}\equiv\tau_R$.

 Using Eq.~\eqref{equ6} we can express $\pi$ in the following manner
\begin{align}
 \pi =  \left(\frac{\partial \varepsilon}{\partial T}~\dot{T}+\frac{\partial \varepsilon}{\partial \mu}~\dot{\mu}\right)\tau+(\varepsilon+p).
 \label{equ9}
\end{align}
In Eq.~\eqref{equ9} $\dot{T}\equiv dT/d\tau$ and $\dot{\mu}\equiv d\mu/d\tau$. Using the constant relaxation time approximation in the shear sector, Eq.~\eqref{equ7} can be written as
\begin{align}
 \tau_R\dot{\pi} &  = \frac{4}{3}\frac{\eta}{\tau}-\left(1+\beta\frac{\tau_R}{\tau}\right)\pi.
 \label{equ10}
\end{align}
Using Eq.~\eqref{equ9} on the right-hand side of Eq.~\eqref{equ10} we obtain
\begin{align}
 \tau_R \dot{\pi} = & 
 \frac{4}{3}\frac{\eta}{\tau}-\left(1+\beta\frac{\tau_R}{\tau}\right)\bigg[\frac{\partial \varepsilon}{\partial T}\dot{T}\tau\nonumber\\
 & ~~~~~~~~~~~~~~~~~~~~~~~~~~~+\frac{\partial \varepsilon}{\partial \mu}\dot{\mu}\tau+(\varepsilon+p)\bigg].
 \label{equ11}
\end{align}
Furthermore, from Eq.~\eqref{equ9} we get
\begin{align}
 \dot{\pi} = & \frac{\partial^2\varepsilon}{\partial T^2}\dot{T}^2\tau+\frac{\partial\varepsilon}{\partial T}\ddot{T}\tau+\left(\frac{\partial p}{\partial T}+2\frac{\partial \varepsilon}{\partial T}\right)\dot{T}
  +\frac{\partial^2\varepsilon}{\partial \mu^2}\dot{\mu}^2\tau\nonumber\\
  & +\frac{\partial\varepsilon}{\partial \mu}\ddot{\mu}\tau+\left(\frac{\partial p}{\partial \mu}+2\frac{\partial \varepsilon}{\partial \mu}\right)\dot{\mu}+2\frac{\partial^2\varepsilon}{\partial\mu\partial T}\dot{\mu}\dot{T}\tau.
 \label{equ12}
\end{align}
Eq.~\eqref{equ11} allows us to write Eq.~\eqref{equ12} in the following way
\begin{align}
  & \tau_R\frac{\partial\varepsilon}{\partial T}\ddot{T}+\tau_R \frac{\partial^2\varepsilon}{\partial T^2}\dot{T}^2+\dot{T}\bigg[\frac{\tau_R}{\tau}\left(\frac{\partial p}{\partial T}+2\frac{\partial\varepsilon}{\partial T}\right)\nonumber\\
 & +\left(1+\beta\frac{\tau_R}{\tau}\right)\frac{\partial\varepsilon}{\partial T}\bigg]+2\tau_R\frac{\partial^2\varepsilon}{\partial T\partial\mu}\dot{T}\dot{\mu}
 +\tau_R\frac{\partial\varepsilon}{\partial \mu}\ddot{\mu}\nonumber\\
 & +\tau_R \frac{\partial^2\varepsilon}{\partial \mu^2}\dot{\mu}^2 +\dot{\mu}\bigg[\frac{\tau_R}{\tau}\left(\frac{\partial p}{\partial \mu}+2\frac{\partial\varepsilon}{\partial \mu}\right) +\left(1+\beta\frac{\tau_R}{\tau}\right)\frac{\partial\varepsilon}{\partial \mu}\bigg]\nonumber\\
 & 
 ~~~~~~~~~~~~~~~~~~~+
  \bigg[\left(1+\beta\frac{\tau_R}{\tau}\right)\frac{\varepsilon+p}{\tau}-\frac{4}{3}\frac{\eta}{\tau^2}\bigg] = 0.
 \label{equ13new}
\end{align}
Equation~\eqref{equ13new} is a second-order differential equation which is derived from the two first-order differential equations (namely, Eqs.~\eqref{equ6} and~\eqref{equ7}). The evolution equation of the number density ($n$), i.e., Eq.~\eqref{equ8} remains a first-order differential equation. 

\subsection{FOCS approach}

In the FOCS approach with the Bjorken-flow geometry, the evolution equations of the energy density ($\mathcal{E}$) and the number density ($\mathcal{N}$) in a general frame become~\cite{Kovtun:2019hdm}
\begin{align}
    & \frac{d\mathcal{E}}{d\tau}+\frac{\mathcal{E}+\mathcal{P}}{\tau}-\frac{4}{3}\frac{\eta}{\tau^2}=0,\label{equnew14}\\
    & \frac{d\mathcal{N}}{d\tau}+\frac{\mathcal{N}}{\tau}=0\label{equnew15}.
\end{align}
Here one considers the following constitutive relations for a charged fluid~\cite{Kovtun:2019hdm}:
\begin{align}
\mathcal{E}& =\varepsilon+\varepsilon_1\frac{\dot{T}}{T}+\varepsilon_2\frac{1}{\tau}+\varepsilon_3\frac{d}{d\tau}\left(\frac{\mu}{T}\right),
\label{equnew16}\\
\mathcal{P}& =p+\pi_1\frac{\dot{T}}{T}+\pi_2\frac{1}{\tau}+\pi_3\frac{d}{d\tau}\left(\frac{\mu}{T}\right),
\label{equnew17}\\
\mathcal{N}& =n+\nu_1\frac{\dot{T}}{T}+\nu_2\frac{1}{\tau}+\nu_3\frac{d}{d\tau}\left(\frac{\mu}{T}\right). 
\label{equnew18}
\end{align}
We note that the dimension of the coefficients $\varepsilon_{i}$  and $\pi_i$ in natural units is GeV$^3$, which is different from the dimension of energy density ($\varepsilon$) or pressure ($p$).
Further the dimension of the coefficients $\nu_i$ in natural units is GeV$^2$  which is different from the dimension of number density ($n$).
The coefficients $\varepsilon_{i}$, $\pi_{i}$ and $\nu_i$ can be interpreted as regulators of the theory. These regulators result in additional non-hydrodynamic modes in the sound and shear channels. Also these regulators are relevant for nonlinear causality, linear stability, and existence and uniqueness of solutions. 

For the boost-invariant Bjorken flow, the heat flow and the baryon diffusion currents identically vanish \footnote{In the FOCS approaches up to a first-order derivative expansion of the hydrodynamic variables the heat flow can be expressed as~\cite{Kovtun:2019hdm},
$\mathcal{Q}^{\mu}\equiv -\Delta_{\mu \alpha} u_{\beta} T^{\alpha \beta} =\theta_{1} \dot{u}^{\mu}+\theta_{2} / T \Delta^{\mu \lambda} \partial_{\lambda} T+\theta_{3} \Delta^{\mu \lambda} \partial_{\lambda}(\mu / T)$.   Similarly the baryon diffusion current can be written as,
$\mathcal{J}^{\mu} \equiv \Delta_{\mu \alpha} J^{\alpha} =\gamma_{1} \dot{u}^{\mu}+\gamma_{2} / T \Delta^{\mu \lambda} \partial_{\lambda} T+\gamma_{3} \Delta^{\mu \lambda} \partial_{\lambda}(\mu / T)$. For Bjorken flow $\dot{u}^{\mu}$, while $\Delta^{\mu \lambda} \partial_{\lambda} T$ and $\Delta^{\mu \lambda} \partial_{\lambda}(\mu / T)$ vanish identically. Therefore coefficients $\theta_i$ and $\gamma_i$ do not contribute to the FOCS equations for Bjorken flow.}.
Using the constitutive relations for the energy density and the pressure (as given by Eqs.~\eqref{equnew16} and \eqref{equnew17}, respectively) in Eq.~\eqref{equnew14}, the FOCS equation for the conservation of the energy-momentum tensor becomes
\begin{align}
     &  \bigg(\frac{\varepsilon_1}{T}-\frac{\varepsilon_3\mu}{T^2}\bigg)\ddot{T}+\bigg(\frac{\partial\varepsilon_1}{\partial T}\frac{1}{T}-\frac{\varepsilon_1}{T^2}-\frac{\partial\varepsilon_3}{\partial T}\frac{\mu}{T^2}+2 \varepsilon_3\frac{\mu}{T^3}\bigg)\dot{T}^2\nonumber\\
    & +\bigg(\frac{\partial\varepsilon_1}{\partial \mu}\frac{1}{T}+\frac{\partial\varepsilon_3}{\partial T}\frac{1}{T}-\frac{\varepsilon_3}{T^2}-\frac{\partial\varepsilon_3}{\partial\mu}\frac{\mu}{T^2}-\frac{\varepsilon_3}{T^2}\bigg)\dot{T}\dot{\mu}\nonumber\\
    & +\bigg(\frac{\varepsilon_3}{T}\bigg)\ddot{\mu}+\bigg(\frac{\partial\varepsilon_3}{\partial\mu}\frac{1}{T}\bigg)\dot{\mu}^2+\bigg(\frac{\partial\varepsilon}{\partial T}+\frac{1}{\tau}\frac{\partial\varepsilon_2}{\partial T}\bigg)\dot{T}\nonumber\\
    & +\bigg(\frac{\partial\varepsilon}{\partial \mu}+\frac{1}{\tau}\frac{\partial\varepsilon_2}{\partial \mu}\bigg)\dot{\mu}+\frac{\varepsilon+p}{\tau}+\frac{\varepsilon_1+\pi_1}{\tau}\frac{\dot{T}}{T}+\frac{\pi_2}{\tau^2}
    \nonumber\\
    & +\frac{\varepsilon_3+\pi_3}{\tau}\frac{\dot{\mu}}{T}-\frac{\varepsilon_3+\pi_3}{\tau}\frac{\mu}{T^2}\dot{T}-\frac{4}{3}\frac{\eta}{\tau^2}=0.
    \label{equ20new}
\end{align}
Moreover, using the constitutive relation  given by Eq.~\eqref{equnew18} the FOCS equation for the conservation of the number current can be written as
\begin{align}
&\bigg(\frac{\nu_1}{T}-\frac{\nu_3\mu}{T^2}\bigg)\ddot{T}+\bigg(\frac{\partial\nu_1}{\partial T}\frac{1}{T}-\frac{\nu_1}{T^2}-\frac{\partial\nu_3}{\partial T}\frac{\mu}{T^2}+2 \nu_3\frac{\mu}{T^3}\bigg)\dot{T}^2\nonumber\\
    &+\bigg(\frac{\partial\nu_1}{\partial \mu}\frac{1}{T}+\frac{\partial\nu_3}{\partial T}\frac{1}{T}-\frac{\nu_3}{T^2}-\frac{\partial\nu_3}{\partial\mu}\frac{\mu}{T^2}-\frac{\nu_3}{T^2}\bigg)\dot{T}\dot{\mu}\nonumber\\
    & +\bigg(\frac{\nu_3}{T}\bigg)\ddot{\mu}+\bigg(\frac{\partial\nu_3}{\partial\mu}\frac{1}{T}\bigg)\dot{\mu}^2+\bigg(\frac{\partial n}{\partial T}+\frac{1}{\tau}\frac{\partial\nu_2}{\partial T}\bigg)\dot{T}\nonumber\\
    & +\bigg(\frac{\partial n}{\partial \mu}+\frac{1}{\tau}\frac{\partial\nu_2}{\partial \mu}\bigg)\dot{\mu}+\frac{n}{\tau}+\frac{\nu_1}{\tau}\frac{\dot{T}}{T}+\frac{\nu_3}{\tau}\frac{\dot{\mu}}{T}-\frac{\nu_3}{\tau}\frac{\mu}{T^2}\dot{T}=0.
    \label{equ21new}
\end{align}

\section{Matching between IS and FOCS theory }
\label{sec3}

Equations~\eqref{equ20new} and \eqref{equ21new} have the structure of the Riccati equation similar to Eq.~\eqref{equ13new}. On the other hand, Eq.~\eqref{equ8} is a first-order equation which can be considered as a special case of the Riccati equation. Equations~\eqref{equ13new} and~\eqref{equ8} in the IS theory can be compared with the corresponding FOCS equations, i.e., Eqs.~\eqref{equ20new} and~\eqref{equ21new} to check if a direct correspondence between the IS and FOCS theory can be constructed. 

If we compare the coefficients multiplying $\ddot{T}$ in Eqs.~\eqref{equ8} and~\eqref{equ21new} we get
 \begin{align}
 \frac{\nu_1}{T}-\nu_3\frac{\mu}{T^2}=0,
 \label{equnu1}
 \end{align}
 where $\nu_3$ can be obtained by comparing the coefficients of $\ddot{\mu}$ in Eqs.~\eqref{equ8} and~\eqref{equ21new}, which gives
 \begin{align}
\nu_3=0.
\label{equ33new}
 \end{align}
 This also directly implies $\nu_1=0$. Therefore, any possible correspondence between the IS and FOCS theory will work in a general frame only  with the conditions $\nu_1=\nu_3=0$. Furthermore, by  comparing the first-order derivative terms and derivative independent terms in Eqs.~\eqref{equ8} and~\eqref{equ21new} we get
 \begin{align}
 & \bigg(\frac{\partial n}{\partial T}\dot{T}+\frac{\partial n}{\partial \mu}\dot{\mu}\bigg)+\frac{1}{\tau}\bigg(\frac{\partial \nu_2}{\partial T}\dot{T}+\frac{\partial \nu_2}{\partial \mu}\dot{\mu}\bigg)+\frac{n}{\tau} = \frac{d n}{d \tau}+\frac{n}{\tau}.
 \end{align}
 Using the above equation we obtain, 
\begin{align}
  \frac{d \nu_2}{d \tau} = 0.
 \label{equ34new}
 \end{align}
 Hence, it turns out that $\nu_2$ is $\tau$ independent. Some comments about $\nu_1=0$ and $\nu_3=0$ are in order here. We observe that even in the presence of a dissipative current ($n^{\mu}$) the conservation of the charge current ($\partial_{\mu}N^{\mu}=0$) in the IS theory for the Bjorken flow is reduced to a first-order differential equation (see Eq.\eqref{equ8}). On the other hand, for a general constitutive relation the charge current conservation equation, $\partial_{\mu}J^{\mu}=0$ in the FOCS theory for the Bjorken flow is a second-order differential equation (see Eq.~\eqref{equ21new})~\cite{Kovtun:2019hdm}. Therefore, when we compare a first-order differential equation with a second-order differential equation, it is natural that the coefficients of the second-order terms should vanish in order to have a one-to-one correspondence between these equations. 
 
 For the Bjorken flow in the FOCS theory with $\nu_1=\nu_3=0$ and $\nu_2$ independent of $\tau$, the evolution of the number density (i.e., Eq.~\eqref{equnew15}) is not affected by the dissipative quantities and the UV regulators do not enter into this equation. Note that in the Landau frame as well as in the Eckart frame, the evolution of the number density $(n)$ is not affected by the dissipative fluxes for the Bjorken flow. In the Eckart frame $n^{\mu}$ is by definition zero and in the Landau frame the $n^{\mu}$ is given by its Navier-Stokes limit $n^{\mu}=\kappa_n\Delta^{\mu\nu}\partial_{\nu}(\mu/T)$ which vanishes identically for the Bjorken flow.   

The evolution of the energy density ($\varepsilon$) in the IS theory is also first order in nature but, interestingly, the dissipative part of the energy-momentum tensor ($\pi^{\mu\nu}$) also enters the energy density evolution equation. Therefore, the evolution of the energy density along with the evolution equation of the dissipative flux gives rise to a second order equation (Eq.~\eqref{equ13new}). 
 
Comparing the coefficients multiplying the terms containing $\ddot{T}$ in Eqs.~\eqref{equ13new} and~\eqref{equ20new} we get
\begin{align}
\varepsilon_1=\tau_R T\frac{\partial\varepsilon}{\partial T}+\frac{\mu}{T}\varepsilon_3,
\label{equ20}
\end{align}
so the coefficient $\varepsilon_3$ can be obtained in terms of the relaxation time in the IS theory by comparing the coefficients of $\ddot{\mu}$ in Eqs.~\eqref{equ13new} and \eqref{equ20new}
\begin{align}
\varepsilon_3=\tau_RT\frac{\partial\varepsilon}{\partial\mu}.
\label{equ21}
\end{align}
The coefficient $\varepsilon_3$ given by Eq.~\eqref{equ21} allows us to write $\varepsilon_1$ in the following manner
\begin{align}
\varepsilon_1=\tau_R T\frac{\partial\varepsilon}{\partial T}+\tau_R\mu\frac{\partial\varepsilon}{\partial\mu}.
\label{equ22}
\end{align}
Subsequently, comparing the coefficients of $\dot{T}^2$, $\dot{\mu}^2$ and $\dot{T}\dot{\mu}$ in Eqs.~\eqref{equ13new} and~\eqref{equ20new} we get:
\begin{align}
& \bigg(\frac{\partial\varepsilon_1}{\partial T}\frac{1}{T}-\frac{\varepsilon_1}{T^2}-\frac{\partial\varepsilon_3}{\partial T}\frac{\mu}{T^2}+2 \varepsilon_3\frac{\mu}{T^3}\bigg)=\tau_R \frac{\partial^2\varepsilon}{\partial T^2}
\label{equ23},\\
& \frac{\partial\varepsilon_3}{\partial\mu}\frac{1}{T}=\tau_R\frac{\partial^2\varepsilon}{\partial\mu^2}
\label{equ24},\\
& \bigg(\frac{\partial\varepsilon_1}{\partial \mu}\frac{1}{T}+\frac{\partial\varepsilon_3}{\partial T}\frac{1}{T}-\frac{\varepsilon_3}{T^2}\nonumber\\
&~~~~~~~~~~~~~~~~~~~~~~~-\frac{\partial\varepsilon_3}{\partial\mu}\frac{\mu}{T^2}-\frac{\varepsilon_3}{T^2}\bigg) = 2 \tau_R \frac{\partial^2\varepsilon}{\partial T \partial\mu},
\label{equ25}
\end{align}
respectively. It can be easily shown that $\varepsilon_1$ and $\varepsilon_3$ given by Eqs.~\eqref{equ22} and~\eqref{equ21} also satisfy Eqs.~\eqref{equ23}, \eqref{equ24} and~\eqref{equ25}. Therefore, Eqs.~\eqref{equ23}, \eqref{equ24} and~\eqref{equ25} are redundant.  

Comparing the terms with  $\dot{T}$ and $\dot{\mu}$ in Eqs.~\eqref{equ13new} and~\eqref{equ20new} we get
 \begin{align}
 & \bigg(\frac{\partial\varepsilon}{\partial T}+\frac{1}{\tau}\frac{\partial\varepsilon_2}{\partial T}\bigg)\dot{T}+\bigg(\frac{\partial\varepsilon}{\partial \mu}+\frac{1}{\tau}\frac{\partial\varepsilon_2}{\partial \mu}\bigg)\dot{\mu}+\frac{\varepsilon_1+\pi_1}{\tau}\frac{\dot{T}}{T}\nonumber\\
 &+\frac{\varepsilon_3+\pi_3}{\tau}\frac{\dot{\mu}}{T}-\frac{\varepsilon_3+\pi_3}{\tau}\frac{\mu}{T^2}\dot{T} \nonumber\\
 &=\dot{T}\bigg[\frac{\tau_R}{\tau}\left(\frac{\partial p}{\partial T}+2\frac{\partial\varepsilon}{\partial T}\right)+\left(1+\beta\frac{\tau_R}{\tau}\right)\frac{\partial\varepsilon}{\partial T}\bigg]\nonumber\\
 & +\dot{\mu}\bigg[\frac{\tau_R}{\tau}\left(\frac{\partial p}{\partial \mu}+2\frac{\partial\varepsilon}{\partial \mu}\right)+\left(1+\beta\frac{\tau_R}{\tau}\right)\frac{\partial\varepsilon}{\partial \mu}\bigg].
 \label{equ26}
 \end{align}
 Equation~\eqref{equ26} can be solved for $\varepsilon_2$, once we know $\pi_1$, $\pi_3$ in terms of the IS parameters. Similarly, by comparing derivative independent terms in Eqs.~\eqref{equ13new} and~\eqref{equ20new} we get
 \begin{align}
 & \frac{\pi_2}{\tau^2}+\frac{\varepsilon+p}{\tau}-\frac{4}{3}\frac{\eta}{\tau^2}=\bigg(1+\beta\frac{\tau_R}{\tau}\bigg)\frac{\varepsilon+p}{\tau}-\frac{4}{3}\frac{\eta}{\tau^2}. \nonumber\\
 \end{align}
 The above equation can be used to determine $\pi_2$,
 \begin{align}
 \pi_2 = \beta \tau_R(\varepsilon+p).
 \label{equ28}
 \end{align}
 
The coefficient of the bulk viscosity in the FOCS approach for a charged fluid can be given as \cite{Hoult:2020eho}
\begin{align}
\zeta = & \bigg(\bigg(\frac{\partial p}{\partial\varepsilon}\bigg)_n\pi_1-\pi_2\bigg)+\bigg(\frac{\partial p}{\partial\varepsilon}\bigg)_n\bigg(\varepsilon_2-\bigg(\frac{\partial p}{\partial\varepsilon}\bigg)_n\varepsilon_1\bigg)\nonumber\\
& +\frac{1}{T}\bigg(\frac{\partial p}{\partial n}\bigg)_{\varepsilon}\bigg(\pi_3-\bigg(\frac{\partial p}{\partial\varepsilon}\bigg)_n\varepsilon_3\bigg)\nonumber\\
& +\bigg(\frac{\partial p}{\partial n}\bigg)_{\varepsilon}\bigg(\nu_2-\bigg(\frac{\partial p}{\partial\varepsilon}\bigg)_n\nu_1\bigg)-\frac{1}{T}\bigg(\frac{\partial p}{\partial n}\bigg)_{\varepsilon}^2\nu_3. 
\label{equ29}
\end{align}
 For an ideal gas equation of state (massless particles)~\cite{Hosoya:1983xm}, 
 \begin{align}
 \bigg(\frac{\partial p}{\partial\varepsilon}\bigg)_n= \frac{1}{3}, ~~~\bigg(\frac{\partial p}{\partial n}\bigg)_{\varepsilon}=0, ~~~\zeta = 0.
 \label{equ30}
 \end{align}
 Using Eqs.~\eqref{equ29} and~\eqref{equ30} we obtain
 \begin{align}
 \pi_1 = 3\pi_2-(\varepsilon_2-\varepsilon_1/3).
 \label{equ31}
 \end{align}
 
 Using Eqs.~\eqref{equ22} and~\eqref{equ28} in Eq.~\eqref{equ31} we can express $\pi_1$ in terms of $\varepsilon_2$ and other thermodynamic quantities. Using the expression of $\pi_1$ in terms of $\varepsilon_2$ in Eq.~\eqref{equ26} we can obtain $\varepsilon_2$ provided we also know $\pi_3$. Till now we have not discussed how to evaluate $\pi_3$, with the other FOCS coefficients it can be obtained uniquely using the condition that for an ideal gas equation of state the trace of the energy-momentum tensor vanishes in the IS theory.
 
\subsection{Traceless condition of the energy-momentum tensor}

As we have already discussed, to calculate $\varepsilon_2$ we also need to know $\pi_3$. For the massless case, the IS theory we are considering has vanishing trace of the energy-momentum tensor. If we also demand the trace of the energy-momentum tensor to vanish for an ideal gas equation of state in the FOCS theory then we get the following conditions~\cite{Kovtun:2019hdm}
\begin{align}
\pi_i= \frac{1}{3}\varepsilon_i,~~i=1,2,3.
\end{align}
Therefore, 
\begin{align}
 \pi_3=\frac{1}{3}\varepsilon_3=\frac{1}{3}\tau_R T \frac{\partial\varepsilon}{\partial\mu}
\end{align}
and
\begin{align}
\varepsilon_2 = 3 \pi_2 = 3\beta \tau_R (\varepsilon+p).
\label{equ39}
\end{align}
It can be shown that for an ideal gas equation of state the coefficient $\varepsilon_2$, as given by Eq.~\eqref{equ39}, does not satisfy Eq.~\eqref{equ26} for any arbitrary value of the IS coefficient $\beta$. In other words, using Eqs.~\eqref{equ39} and~\eqref{equ26} we can get the IS parameter $\beta$ for which  $\varepsilon_2$ can be uniquely determined. This can be done either by comparing the coefficients of $\dot{T}$ or $\dot{\mu}$ from both sides of Eq.~\eqref{equ26}. Therefore, by comparing the coefficients of $\dot{T}$ from both sides of Eq.~\eqref{equ26} we get:
\begin{align}
& \bigg(\frac{\partial\varepsilon}{\partial T}+\frac{1}{\tau}\frac{\partial\varepsilon_2}{\partial T}\bigg)\dot{T}+\frac{\varepsilon_1+\pi_1}{\tau}\frac{\dot{T}}{T}-\frac{\varepsilon_3+\pi_3}{\tau}\frac{\mu}{T^2}\dot{T}
\nonumber\\
& = \dot{T}\bigg[\frac{\tau_R}{\tau}\left(\frac{\partial p}{\partial T}+2\frac{\partial\varepsilon}{\partial T}\right)+\left(1+\beta\frac{\tau_R}{\tau}\right)\frac{\partial\varepsilon}{\partial T}\bigg]\nonumber\\
\implies & \bigg(\frac{1}{\tau}\frac{\partial\varepsilon_2}{\partial T}\bigg)+ \frac{4}{3}\frac{\varepsilon_1}{\tau T}-\frac{4}{3}\frac{\varepsilon_3}{\tau}\frac{\mu}{T^2} \nonumber\\
& ~~~~~~~~~~~~~~~~= \bigg[\frac{\tau_R}{\tau}\left(\frac{\partial p}{\partial T}+2\frac{\partial\varepsilon}{\partial T}\right)+\beta\frac{\tau_R}{\tau}\frac{\partial\varepsilon}{\partial T}\bigg]\nonumber\\
\implies & 3 \beta \tau_R \bigg(\frac{\partial\varepsilon}{\partial T}+\frac{\partial p}{\partial T}\bigg)+\frac{4}{3}\bigg(\tau_R T \frac{\partial\varepsilon}{\partial T}+\tau_R\mu\frac{\partial \varepsilon}{\partial \mu}\bigg)\frac{1}{T}\nonumber\\
& -\frac{4}{3}\tau_R T \frac{\partial\varepsilon}{\partial\mu}\frac{\mu}{T^2} = \bigg[\tau_R\left(\frac{\partial p}{\partial T}+2\frac{\partial\varepsilon}{\partial T}\right)+\beta \tau_R\frac{\partial\varepsilon}{\partial T}\bigg],\nonumber\\
\implies & (3\beta+1)\tau_R\frac{4}{3}\frac{\partial\varepsilon}{\partial T} = \tau_R\bigg(\frac{7}{3}+\beta\bigg)\frac{\partial\varepsilon}{\partial T} \nonumber\\
\implies & \beta = \frac{1}{3}.
\label{equ38new}
\end{align}
In the second line of Eq.~\eqref{equ38new} we have used the relations $\pi_1=(1/3)\varepsilon_1$ and $\pi_3=(1/3)\varepsilon_3$. We have also used Eqs.~\eqref{equ21}, \eqref{equ22}, and~ \eqref{equ39}, and the ideal gas equation of state to get Eq.~\eqref{equ38new}. Similarly, by comparing the coefficients of $\dot{\mu}$ from both sides of Eq.~\eqref{equ26} it can be shown that $\beta=1/3$. Therefore, only for $\beta = 1/3$ the coefficient $\varepsilon_2$ along with the other regulators can be uniquely determined. In this case we obtain
\begin{align}
\varepsilon_2 = \tau_R(\varepsilon+p)=\frac{4}{3}\tau_R\varepsilon.
\end{align}
As a consequence, for an ideal gas EOS and $\beta=1/3$ we obtain the following relations:
\begin{align}
& \varepsilon_1 = \tau_R T \frac{\partial\varepsilon}{\partial T} +\tau_R \mu \frac{\partial\varepsilon}{\partial \mu}= 3 \tau_R T \frac{\partial p}{\partial T} +3 \tau_R \mu \frac{\partial p}{\partial \mu}\nonumber\\
& ~~~~=3\tau_R Ts+3\tau_R \mu n =3\tau_R (\varepsilon+p)=4\tau_R\varepsilon, \nonumber\\
\label{equ41new}\\
& \varepsilon_2 = \tau_R(\varepsilon+p)=\frac{4}{3}\tau_R\varepsilon,\label{equ42new}\\
& \varepsilon_3 = \tau_R T \frac{\partial\varepsilon}{\partial \mu} = 3 \tau_R T \frac{\partial p}{\partial \mu} = 3 \tau_R T n,\label{equ43new}\\
& \pi_1 = \tau_R(\varepsilon+p)=\frac{4}{3}\tau_R\varepsilon,\label{equ44new}\\
& \pi_2 = \frac{1}{3}\tau_R(\varepsilon+p)=\frac{4}{9}\tau_R\varepsilon,\label{equ45new}\\
& \pi_3 = \tau_R T n.\label{equ46new}
\end{align}
Eqs.~\eqref{equ41new}--\eqref{equ46new} allow us to express the FOCS regulator sector in terms of the IS relaxation time. For vanishing baryon chemical potential ($\mu=0$ or $n=0$) Eqs.~\eqref{equ41new}--\eqref{equ46new}  exactly agree with our previous results presented in Ref.~\cite{Das:2020fnr}. Note that for $\mu=0$ the coefficients $\varepsilon_3$ and $\pi_3$ are not present in the FOCS constitutive relations.

\section{Causality and Stability}
\label{caisality_stability}

 We have shown that hydrodynamic equations in the FOCS approach in a frame with $\nu_1=\nu_3=0$ can be exactly mapped into the hydrodynamic equations of Israel-Stewart (IS) theory for the boost-invariant flow at finite baryon density with an ideal gas equation of state. Therefore, we can use the knowledge of stability and causality in the FOCS sector to learn about the stability and causality of the IS theory considered here. Causality and stability in the FOCS approach for a charged fluid have been discussed in Ref.~\cite{Hoult:2020eho}. The stability and causality of the shear modes requires $\theta_1>\eta>0$. Note that $\theta_1$ is the transport coefficient associated with the heat flow $\mathcal{Q}^{\mu}$ in the FOCS approach. For the Bjorken flow $\mathcal{Q}^{\mu}$ identically vanishes. Therefore, the coefficient $\theta_1$ cannot be constrained in this case. This means that the causality and stability conditions for the shear modes can be satisfied. 
 
 What remains to check is the causality and stability for the sound channels. Unfortunately, the causality and stability of sound channels in the most general frame in the FOCS approach are still not completely explored. Only for some specific frame choice with $\varepsilon_3=\pi_3=\theta_3=0$ the causality and stability in a Lorentz boosted frame in the FOCS sector has been discussed recently in Ref.~\cite{Hoult:2020eho}. In the present investigation, we have found from the constructed correspondence that $\varepsilon_3$ and $\pi_3$ do not vanish. Therefore, we cannot make conclusive statements regarding the causality and stability in a Lorentz boosted frame for the case with $\varepsilon_3\neq 0$ and $\pi_3\neq 0$. However, we can make some important comments about the stability in the fluid rest frame for the sound modes. One of the necessary conditions for the stability of equilibrium in the fluid rest frame is $\varepsilon_1\nu_3-\varepsilon_3\nu_1>0$~\cite{Hoult:2020eho}. Following Eqs.~\eqref{equ42new} and~\eqref{equ44new} we have $\varepsilon_1\geq 0$ and $\varepsilon_3\geq 0$. Therefore, for $\nu_1=\nu_3=0$, $\varepsilon_1\geq 0$, and $\varepsilon_3\geq 0$ at least one of the stability conditions in the fluid rest frame cannot be satisfied. Hence, the FOCS theory or equivalently the IS theory is unstable for $\nu_1=\nu_3=0$, $\varepsilon_1\geq 0$, and $\varepsilon_3\geq 0$. However, this instability does not guarantee that the hydrodynamic theory is acausal. We should also mention that recently in Ref.~\cite{Bemfica:2020zjp} a general relativistic formulation of causal and stable first order hydrodynamics has been discussed for finite baryon density using a hydrodynamic frame which is closer to Eckart frame where the flow velocity is defined using the baryon current. In such a formulation, the conserved charge current is free of any dissipative part which is different from the formalism discussed in Ref.~\cite{Hoult:2020eho}. Although the hydrodynamic formalism as described in Ref.~\cite{Hoult:2020eho} applies to baryon rich matter, e.g. neutron star mergers or in low energy heavy-ion collisions, a boost invariant flow is not physically appealing in such a situation.
 
 It is important to note that in a hydrodynamic theory causality and stability are intimately related. It has been argued in Refs.~\cite{Pu:2009fj, Denicol:2008ha} that in a Lorentz boosted frame hydrodynamic theory with nonvanishing bulk and shear viscosities develops instability if the theory is acausal. However the opposite may not be true, i.e., a causal theory may develop instability. Till now we have made comments on the stability and causality of the IS theory for Bjorken flow using our knowledge of FOCS theory. One can also use the knowledge of IS theory to shed light on the FOCS theory using the correspondence between them. But causality and stability in the IS theory for a boost-invariant Bjorken flow at finite baryon chemical potential are still not explored extensively. Furthermore, a detailed investigation is required to understand the causality and stability of the IS theory at finite baryon chemical potential. Only recently the nonlinear causality and stability of an uncharged fluid have been explored in Ref.~\cite{Bemfica:2020xym}.      

 \section{summary}
 \label{summary}
In the present study we have found an exact correspondence between second-order Israel-Stewart (IS) hydrodynamic theory and first-order causal and stable (FOCS) hydrodynamics for the boost-invariant flow with nonvanishing baryon density for massless particles. The crucial assumption that allowed for this correspondence is the constant relaxation time in the shear sector of the IS theory. Explicit expressions for the temperature and chemical potential dependent regulators of the FOCS theory have been given for an ideal gas equation of state. The stability and causality criteria known in the FOCS approach have been applied to the IS theory. We have found that the conditions for which the correspondence between the IS and FOCS theory exists, does not satisfy at least one of the necessary conditions for the stability of the theory. 

\medskip
{\bf Acknowledgements:} We thank J. Noronha for many illuminating comments. We would also like to thank A. Jaiswal for very useful discussions on the Israel-Stewart theory at finite baryon chemical potential. This work was supported in part by the Polish National Science Center Grants No.~2016/23/B/ST2/00717 and No.~2018/30/E/ST2/00432.
\medskip

\bibliographystyle{utphys} 


\begin{thebibliography}{10}

\bibitem{Florkowski:2010zz}
W.~Florkowski, {\em {Phenomenology of Ultra-Relativistic Heavy-Ion
  Collisions}}.
\newblock
2010.
\newblock

\bibitem{Gale:2013da}
C.~Gale, S.~Jeon, and B.~Schenke, ``{Hydrodynamic Modeling of Heavy-Ion
  Collisions},'' \href{http://dx.doi.org/10.1142/S0217751X13400113}{{\em Int.
  J. Mod. Phys.} {\bf A28} (2013)  1340011},
\href{http://arxiv.org/abs/1301.5893}{{\tt arXiv:1301.5893 [nucl-th]}}.

\bibitem{Jeon:2015dfa}
S.~Jeon and U.~Heinz, ``{Introduction to Hydrodynamics},''
  \href{http://dx.doi.org/10.1142/S0218301315300106}{{\em Int. J. Mod. Phys.}
  {\bf E24} (2015) no.~10, 1530010},
\href{http://arxiv.org/abs/1503.03931}{{\tt arXiv:1503.03931 [hep-ph]}}.

\bibitem{Jaiswal:2016hex}
A.~Jaiswal and V.~Roy, ``{Relativistic hydrodynamics in heavy-ion collisions:
  general aspects and recent developments},''
  \href{http://dx.doi.org/10.1155/2016/9623034}{{\em Adv. High Energy Phys.}
  {\bf 2016} (2016)  9623034},
\href{http://arxiv.org/abs/1605.08694}{{\tt arXiv:1605.08694 [nucl-th]}}.

\bibitem{Busza:2018rrf}
W.~Busza, K.~Rajagopal, and W.~van~der Schee, ``{Heavy Ion Collisions: The Big
  Picture, and the Big Questions},''
  \href{http://dx.doi.org/10.1146/annurev-nucl-101917-020852}{{\em Ann. Rev.
  Nucl. Part. Sci.} {\bf 68} (2018)  339--376},
\href{http://arxiv.org/abs/1802.04801}{{\tt arXiv:1802.04801 [hep-ph]}}.

\bibitem{Romatschke:2017ejr}
P.~Romatschke and U.~Romatschke,
  \href{http://dx.doi.org/10.1017/9781108651998}{{\em {Relativistic Fluid
  Dynamics In and Out of Equilibrium}}}.
\newblock Cambridge Monographs on Mathematical Physics. Cambridge University
  Press, 2019.
\newblock
\href{http://arxiv.org/abs/1712.05815}{{\tt arXiv:1712.05815 [nucl-th]}}.
\newblock

\bibitem{Florkowski:2017olj}
W.~Florkowski, M.~P. Heller, and M.~Spalinski, ``{New theories of relativistic
  hydrodynamics in the LHC era},''
  \href{http://dx.doi.org/10.1088/1361-6633/aaa091}{{\em Rept. Prog. Phys.}
  {\bf 81} (2018) no.~4, 046001},
\href{http://arxiv.org/abs/1707.02282}{{\tt arXiv:1707.02282 [hep-ph]}}.

\bibitem{Heller:2015dha}
M.~P. Heller and M.~Spalinski, ``{Hydrodynamics Beyond the Gradient Expansion:
  Resurgence and Resummation},''
  \href{http://dx.doi.org/10.1103/PhysRevLett.115.072501}{{\em Phys. Rev.
  Lett.} {\bf 115} (2015) no.~7, 072501},
\href{http://arxiv.org/abs/1503.07514}{{\tt arXiv:1503.07514 [hep-th]}}.

\bibitem{Romatschke:2017vte}
P.~Romatschke, ``{Relativistic Fluid Dynamics Far From Local Equilibrium},''
  \href{http://dx.doi.org/10.1103/PhysRevLett.120.012301}{{\em Phys. Rev.
  Lett.} {\bf 120} (2018) no.~1, 012301},
\href{http://arxiv.org/abs/1704.08699}{{\tt arXiv:1704.08699 [hep-th]}}.

\bibitem{Strickland:2017kux}
M.~Strickland, J.~Noronha, and G.~Denicol, ``{Anisotropic nonequilibrium
  hydrodynamic attractor},''
  \href{http://dx.doi.org/10.1103/PhysRevD.97.036020}{{\em Phys. Rev.} {\bf
  D97} (2018) no.~3, 036020},
\href{http://arxiv.org/abs/1709.06644}{{\tt arXiv:1709.06644 [nucl-th]}}.

\bibitem{Strickland:2018ayk}
M.~Strickland, ``{The non-equilibrium attractor for kinetic theory in
  relaxation time approximation},''
  \href{http://dx.doi.org/10.1007/JHEP12(2018)128}{{\em JHEP} {\bf 12} (2018)
  128},
\href{http://arxiv.org/abs/1809.01200}{{\tt arXiv:1809.01200 [nucl-th]}}.

\bibitem{Jaiswal:2019cju}
S.~Jaiswal, C.~Chattopadhyay, A.~Jaiswal, S.~Pal, and U.~Heinz, ``{Exact
  solutions and attractors of higher-order viscous fluid dynamics for Bjorken
  flow},'' \href{http://dx.doi.org/10.1103/PhysRevC.100.034901}{{\em Phys.
  Rev.} {\bf C100} (2019) no.~3, 034901},
\href{http://arxiv.org/abs/1907.07965}{{\tt arXiv:1907.07965 [nucl-th]}}.

\bibitem{Giacalone:2019ldn}
G.~Giacalone, A.~Mazeliauskas, and S.~Schlichting, ``{Hydrodynamic attractors,
  initial state energy and particle production in relativistic nuclear
  collisions},'' \href{http://dx.doi.org/10.1103/PhysRevLett.123.262301}{{\em
  Phys. Rev. Lett.} {\bf 123} (2019) no.~26, 262301},
\href{http://arxiv.org/abs/1908.02866}{{\tt arXiv:1908.02866 [hep-ph]}}.

\bibitem{Heller:2013fn}
M.~P. Heller, R.~A. Janik, and P.~Witaszczyk, ``{Hydrodynamic Gradient
  Expansion in Gauge Theory Plasmas},''
  \href{http://dx.doi.org/10.1103/PhysRevLett.110.211602}{{\em Phys. Rev.
  Lett.} {\bf 110} (2013) no.~21, 211602},
\href{http://arxiv.org/abs/1302.0697}{{\tt arXiv:1302.0697 [hep-th]}}.

\bibitem{Denicol:2016bjh}
G.~S. Denicol and J.~Noronha, ``{Divergence of the Chapman-Enskog expansion in
  relativistic kinetic theory},''
\href{http://arxiv.org/abs/1608.07869}{{\tt arXiv:1608.07869 [nucl-th]}}.

\bibitem{Heller:2016rtz}
M.~P. Heller, A.~Kurkela, M.~Spaliński, and V.~Svensson, ``{Hydrodynamization
  in kinetic theory: Transient modes and the gradient expansion},''
  \href{http://dx.doi.org/10.1103/PhysRevD.97.091503}{{\em Phys. Rev.} {\bf
  D97} (2018) no.~9, 091503},
\href{http://arxiv.org/abs/1609.04803}{{\tt arXiv:1609.04803 [nucl-th]}}.

\bibitem{Grozdanov:2019kge}
S.~Grozdanov, P.~K. Kovtun, A.~O. Starinets, and P.~Tadić, ``{Convergence of
  the Gradient Expansion in Hydrodynamics},''
  \href{http://dx.doi.org/10.1103/PhysRevLett.122.251601}{{\em Phys. Rev.
  Lett.} {\bf 122} (2019) no.~25, 251601},
\href{http://arxiv.org/abs/1904.01018}{{\tt arXiv:1904.01018 [hep-th]}}.

\bibitem{Broniowski:2008vp}
W.~Broniowski, M.~Chojnacki, W.~Florkowski, and A.~Kisiel, ``{Uniform
  Description of Soft Observables in Heavy-Ion Collisions at s(NN)**(1/2) = 200
  GeV**2},'' \href{http://dx.doi.org/10.1103/PhysRevLett.101.022301}{{\em Phys.
  Rev. Lett.} {\bf 101} (2008)  022301},
\href{http://arxiv.org/abs/0801.4361}{{\tt arXiv:0801.4361 [nucl-th]}}.

\bibitem{Bozek:2009dw}
P.~Bozek, ``{Bulk and shear viscosities of matter created in relativistic
  heavy-ion collisions},''
  \href{http://dx.doi.org/10.1103/PhysRevC.81.034909}{{\em Phys. Rev.} {\bf
  C81} (2010)  034909},
\href{http://arxiv.org/abs/0911.2397}{{\tt arXiv:0911.2397 [nucl-th]}}.

\bibitem{Noronha-Hostler:2013gga}
J.~Noronha-Hostler, G.~S. Denicol, J.~Noronha, R.~P.~G. Andrade, and F.~Grassi,
  ``{Bulk Viscosity Effects in Event-by-Event Relativistic Hydrodynamics},''
  \href{http://dx.doi.org/10.1103/PhysRevC.88.044916}{{\em Phys. Rev.} {\bf
  C88} (2013) no.~4, 044916},
\href{http://arxiv.org/abs/1305.1981}{{\tt arXiv:1305.1981 [nucl-th]}}.

\bibitem{Bernhard:2019bmu}
J.~E. Bernhard, J.~S. Moreland, and S.~A. Bass, ``{Bayesian estimation of the
  specific shear and bulk viscosity of quark–gluon plasma},''
\href{http://dx.doi.org/10.1038/s41567-019-0611-8}{{\em Nature Phys.} {\bf 15}
  (2019) no.~11, 1113--1117}.

\bibitem{Romatschke:2007mq}
P.~Romatschke and U.~Romatschke, ``{Viscosity Information from Relativistic
  Nuclear Collisions: How Perfect is the Fluid Observed at RHIC?},''
  \href{http://dx.doi.org/10.1103/PhysRevLett.99.172301}{{\em Phys. Rev. Lett.}
  {\bf 99} (2007)  172301},
\href{http://arxiv.org/abs/0706.1522}{{\tt arXiv:0706.1522 [nucl-th]}}.

\bibitem{Gale:2012rq}
C.~Gale, S.~Jeon, B.~Schenke, P.~Tribedy, and R.~Venugopalan, ``{Event-by-event
  anisotropic flow in heavy-ion collisions from combined Yang-Mills and viscous
  fluid dynamics},''
  \href{http://dx.doi.org/10.1103/PhysRevLett.110.012302}{{\em Phys. Rev.
  Lett.} {\bf 110} (2013) no.~1, 012302},
\href{http://arxiv.org/abs/1209.6330}{{\tt arXiv:1209.6330 [nucl-th]}}.

\bibitem{Hiscock:1983zz}
W.~A. Hiscock and L.~Lindblom, ``{Stability and causality in dissipative
  relativistic fluids},''
\href{http://dx.doi.org/10.1016/0003-4916(83)90288-9}{{\em Annals Phys.} {\bf
  151} (1983)  466--496}.

\bibitem{Hiscock:1985zz}
W.~A. Hiscock and L.~Lindblom, ``{Generic instabilities in first-order
  dissipative relativistic fluid theories},''
\href{http://dx.doi.org/10.1103/PhysRevD.31.725}{{\em Phys. Rev.} {\bf D31}
  (1985)  725--733}.

\bibitem{Israel:1976tn}
W.~Israel, ``{Nonstationary irreversible thermodynamics: A Causal relativistic
  theory},''
\href{http://dx.doi.org/10.1016/0003-4916(76)90064-6}{{\em Annals Phys.} {\bf
  100} (1976)  310--331}.

\bibitem{Israel:1979wp}
W.~Israel and J.~M. Stewart, ``{Transient relativistic thermodynamics and
  kinetic theory},''
\href{http://dx.doi.org/10.1016/0003-4916(79)90130-1}{{\em Annals Phys.} {\bf
  118} (1979)  341--372}.

\bibitem{Bemfica:2017wps}
F.~S. Bemfica, M.~M. Disconzi, and J.~Noronha, ``{Causality and existence of
  solutions of relativistic viscous fluid dynamics with gravity},''
  \href{http://dx.doi.org/10.1103/PhysRevD.98.104064}{{\em Phys. Rev.} {\bf
  D98} (2018) no.~10, 104064},
\href{http://arxiv.org/abs/1708.06255}{{\tt arXiv:1708.06255 [gr-qc]}}.

\bibitem{Bemfica:2019knx}
F.~S. Bemfica, M.~M. Disconzi, and J.~Noronha, ``{Nonlinear Causality of
  General First-Order Relativistic Viscous Hydrodynamics},''
  \href{http://dx.doi.org/10.1103/PhysRevD.100.104020}{{\em Phys. Rev.} {\bf
  D100} (2019) no.~10, 104020},
\href{http://arxiv.org/abs/1907.12695}{{\tt arXiv:1907.12695 [gr-qc]}}.

\bibitem{Kovtun:2019hdm}
P.~Kovtun, ``{First-order relativistic hydrodynamics is stable},''
  \href{http://dx.doi.org/10.1007/JHEP10(2019)034}{{\em JHEP} {\bf 10} (2019)
  034},
\href{http://arxiv.org/abs/1907.08191}{{\tt arXiv:1907.08191 [hep-th]}}.

\bibitem{Das:2020fnr}
A.~Das, W.~Florkowski, J.~Noronha, and R.~Ryblewski, ``{Equivalence between
  first-order causal and stable hydrodynamics and Israel-Stewart theory for
  boost-invariant systems with a constant relaxation time},''
  \href{http://dx.doi.org/10.1016/j.physletb.2020.135525}{{\em Phys. Lett. B}
  {\bf 806} (2020)  135525}.

\bibitem{Das:2020gtq}
A.~Das, W.~Florkowski, and R.~Ryblewski, ``{Correspondence between
  Israel-Stewart and first-order casual and stable hydrodynamics for the
  boost-invariant massive case with zero baryon density},''
  \href{http://dx.doi.org/10.1103/PhysRevD.102.031501}{{\em Phys. Rev. D} {\bf
  102} (2020) no.~3, 031501}.

\bibitem{Bemfica:2020xym}
F.~S. Bemfica, M.~M. Disconzi, V.~Hoang, J.~Noronha, and M.~Radosz,
  ``{Nonlinear Constraints on Relativistic Fluids Far From Equilibrium},''
  \href{http://arxiv.org/abs/2005.11632}{{\tt arXiv:2005.11632 [hep-th]}}.

\bibitem{Hoult:2020eho}
R.~E. Hoult and P.~Kovtun, ``{Stable and causal relativistic Navier-Stokes
  equations},'' \href{http://dx.doi.org/10.1007/JHEP06(2020)067}{{\em JHEP}
  {\bf 06} (2020)  067}, \href{http://arxiv.org/abs/2004.04102}{{\tt
  arXiv:2004.04102 [hep-th]}}.

\bibitem{Jaiswal:2013fc}
A.~Jaiswal, R.~S. Bhalerao, and S.~Pal, ``{Complete relativistic second-order
  dissipative hydrodynamics from the entropy principle},''
  \href{http://dx.doi.org/10.1103/PhysRevC.87.021901}{{\em Phys. Rev. C} {\bf
  87} (2013) no.~2, 021901}, \href{http://arxiv.org/abs/1302.0666}{{\tt
  arXiv:1302.0666 [nucl-th]}}.

\bibitem{Jaiswal:2015mxa}
A.~Jaiswal, B.~Friman, and K.~Redlich, ``{Relativistic second-order dissipative
  hydrodynamics at finite chemical potential},''
  \href{http://dx.doi.org/10.1016/j.physletb.2015.11.018}{{\em Phys. Lett. B}
  {\bf 751} (2015)  548--552}, \href{http://arxiv.org/abs/1507.02849}{{\tt
  arXiv:1507.02849 [nucl-th]}}.

\bibitem{Denicol:2017lxn}
G.~S. Denicol and J.~Noronha, ``{Analytical attractor and the divergence of the
  slow-roll expansion in relativistic hydrodynamics},''
  \href{http://dx.doi.org/10.1103/PhysRevD.97.056021}{{\em Phys. Rev.} {\bf
  D97} (2018) no.~5, 056021},
\href{http://arxiv.org/abs/1711.01657}{{\tt arXiv:1711.01657 [nucl-th]}}.

\bibitem{Du:2019obx}
L.~Du and U.~Heinz, ``{(3+1)-dimensional dissipative relativistic fluid
  dynamics at non-zero net baryon density},''
  \href{http://dx.doi.org/10.1016/j.cpc.2019.107090}{{\em Comput. Phys.
  Commun.} {\bf 251} (2020)  107090},
  \href{http://arxiv.org/abs/1906.11181}{{\tt arXiv:1906.11181 [nucl-th]}}.

\bibitem{Hosoya:1983xm}
A.~Hosoya and K.~Kajantie, ``{Transport Coefficients of QCD Matter},''
  \href{http://dx.doi.org/10.1016/0550-3213(85)90499-7}{{\em Nucl. Phys. B}
  {\bf 250} (1985)  666--688}.

\bibitem{Bemfica:2020zjp}
F.~S. Bemfica, M.~M. Disconzi, and J.~Noronha, ``{General-Relativistic Viscous
  Fluid Dynamics},'' \href{http://arxiv.org/abs/2009.11388}{{\tt
  arXiv:2009.11388 [gr-qc]}}.

\bibitem{Pu:2009fj}
S.~Pu, T.~Koide, and D.~H. Rischke, ``{Does stability of relativistic
  dissipative fluid dynamics imply causality?},''
  \href{http://dx.doi.org/10.1103/PhysRevD.81.114039}{{\em Phys. Rev.} {\bf
  D81} (2010)  114039},
\href{http://arxiv.org/abs/0907.3906}{{\tt arXiv:0907.3906 [hep-ph]}}.

\bibitem{Denicol:2008ha}
G.~Denicol, T.~Kodama, T.~Koide, and P.~Mota, ``{Stability and Causality in
  relativistic dissipative hydrodynamics},''
  \href{http://dx.doi.org/10.1088/0954-3899/35/11/115102}{{\em J. Phys. G} {\bf
  35} (2008)  115102}, \href{http://arxiv.org/abs/0807.3120}{{\tt
  arXiv:0807.3120 [hep-ph]}}.

\end{thebibliography}
\providecommand{\href}[2]{#2}\begingroup\raggedright\endgroup
\end{document}